\documentclass[12pt]{article}
\usepackage{graphics}
\begin{document}
\begin{titlepage}
\begin{flushright}
hep-th/0105012 \\
TIT/HEP-466 \\
May,2001
\end{flushright}
\vspace{0.5cm}
\begin{center}
{\Large \bf 
Scattering of Noncommutative $(n,1)$ Solitons
}
\lineskip .75em
\vskip2.5cm
{\large Takeo Araki } and {\large Katsushi Ito}
\vskip 1.5em
{\large\it Department of Physics\\
Tokyo Institute of Technology\\
Oh-okayama, Meguro,Tokyo, 152-8551, Japan}  
\vskip 3.5em
\end{center}
\vskip3cm
\begin{abstract}
We study scattering of noncommutative 
solitons in $2+1$ dimensional scalar field theory.  
In particular,  we investigate a system of two solitons with 
level $n$ and $n'$ (the $(n,n')$-system) in the large
noncommutativity limit.  
We show that the scattering of a general $(n,n')$-system 
occurs at right angles in the case of zero impact parameter.  
We also derive an exact K\"ahler potential and the metric of the 
moduli space of the $(n,1)$-system.  
We examine numerically the $(n,1)$ scattering 
and find that the closest distance for a fixed scattering angle 
is well approximated by a function  
$a+b\sqrt{n}$ where $a$ and $b$ are some numerical constants.  
\end{abstract}
\end{titlepage}
\baselineskip=0.7cm

It has been recognized that noncommutative field theory is interesting
because it has various non-trivial solitons
\cite{NS,GMS,P,JMW,B,HKL}
which include those do not appear in commutative theories.  
The noncommutative soliton in $2+1$ dimensional scalar field theory 
\cite{GMS,Gorsky,Zhou,Solovyov}
is one of important examples.
Noncommutative field theory arises also in open string theory 
with a $B$-field background \cite{Schomerus,SW};  
the effective field theory on D-branes becomes noncommutative.
Noncommutative solitons of such field theory can be identified with 
various D-brane confugurations 
\cite{DMR,Harvey,GN,AGMS,HKL}.   
For example, 
one may use tachyonic noncommutative solitons to construct lower dimensional 
D-branes from unstable higher dimensional D-branes 
in the effective field theoretic description of open string field theory 
\cite{Harvey}.  
In order to study dynamical aspects of D-branes, scattering of
noncommutative solitons \cite{LRU} would play an important role.
Recently,  
a systematic construction of a multi-soliton solution 
was proposed \cite{GHS,HLRU}.  
In these works,  
the level one multi-soliton solutions and finite $\theta$
correction have been explored.  
Here,  a level one soliton is a radially symmetric Gaussian lump 
solution.  

In this letter, we will study scattering of  
noncommutative solitons with higher levels 
in the limit of the large noncommutativity($\theta\rightarrow\infty$).  
In particular, we will consider a system which contains two solitons 
with level $n$ and $n'$, respectively. 
A ``level $n$'' soliton means $n$ coincident level one 
solitons.  
We call this system the $(n,n')$-system.  
We investigate 
the K\"ahler potential and metric of the moduli space of 
the $(n,n')$-system.  
In this work, 
using the expansion around the origin of the moduli space,
we will show that the scattering of the $(n,n')$-system 
occurs at right angles in the case of zero impact parameter. 
We next derive an exact K\"ahler potential and the metric of 
the moduli space of the $(n,1)$-system.  
Using this metric we investigate some global aspects of the moduli space.  
We calculate numerically the scattering angle for the $(n,1)$-system 
as a function of the closest distance 
and find that the closest distance is well approximated 
by a function $a+b\sqrt{n}$ ($a$ and $b$ are some constants) 
for a fixed scattering angle.  

We begin with real scalar field theory 
on $2+1$ dimensional noncommutative spacetime with coordinates $(t,x,y)$.
It has spatial noncommutativity such as 
\begin{equation}
[\hat{x},\hat{y}]=i\theta.  
\end{equation}
The action is 
\begin{equation}
S=-\int dtd^2x \left\{\frac 12 (\partial_\mu\phi)^2+V(\phi)\right\},  
\end{equation}
where fields multiplication are defined by using the star product
\begin{equation}
(\phi_1\star\phi_2)(x,y)
=\left.\exp\left(\frac i2 \theta(\partial_{\xi_1}\partial_{\eta_2}
	-\partial_{\eta_1}\partial_{\xi_2})\right)
\phi_1(\xi_1,\eta_1)\phi_2(\xi_2,\eta_2)\right|_{\xi_1=\xi_2=x,\ 
	\eta_1=\eta_2=y}.
\end{equation}
For simplicity,  we assume that the potential function $V$ has 
only one local minimum at $\phi=\lambda$ other than $\phi=0$,  and $V(0)=0$.  
In the following, we will only consider the case of the large 
noncommutativity limit,  $\theta\rightarrow\infty$. 
Rescaling $x,y\rightarrow \sqrt{\theta}x,\sqrt{\theta}y$,  
the action is dominated by the potential term so that the static 
field equation becomes 
\begin{equation}\label{eom}
{\partial V\over \partial\phi}(\phi)=0.  
\end{equation}
One class of solutions to this equation can be constructed by using
a function which satisfies the condition 
\begin{equation}\label{proj}
(\phi_0\star\phi_0)(x,y)=\phi_0(x,y).  
\end{equation}
A solution to eq.(\ref{eom}) is given by $\lambda \phi_0(x,y)$\cite{GMS}.  
A function $\phi(x,y)$ on the noncommutative plane 
can be mapped to an operator $\Phi(\hat{x},\hat{y})$ acting on 
the Hilbert space ${\cal H}$ of one particle on the line. 
The relation  between $\phi(x,y)$ and $\Phi(\hat{x},\hat{y})$ is given 
by using the Weyl-Moyal correspondence:  
\begin{eqnarray}
\Phi(\hat{x},\hat{y})
=\int{d^2k\over(2\pi)}\tilde{\phi}(k)e^{-i(k_x \hat{x}+k_y\hat{y})},  \\
\tilde{\phi}(k)=\int d^2x \phi(x,y)e^{i(k_x x+k_y y)}.  
\end{eqnarray}
With this correspondence,  the energy of a solution $\phi$ 
can be written as  
\begin{equation}\label{energy}
E[\phi]=\theta\int d^2x V(\phi)
=2\pi\theta{\rm Tr}_{\cal H}(V(\Phi)).  
\end{equation}
This formula tells us that if we find a solution $\Phi$,  
then another solution which has the same energy 
can be obtained by acting a unitary operator $U$ 
on $\Phi$ as $U\Phi U^\dagger$.   

An operator which satisfies 
the condition (\ref{proj}) is a projection operator.  
Thus the most general solution of this class can be written 
using a set of orthogonal projection operators $\{P_i\}$:  
\begin{equation}
\lambda(P_1+P_2+\cdots).  
\end{equation}
Taking $a={1\over\sqrt{2}}(\hat{x}+i\hat{y})$ 
and $a^\dagger={1\over \sqrt{2}}(\hat{x}-i\hat{y})$,  
the Hilbert space can be regarded as 
the Fock space of a harmonic oscillator.  
Any projection operator can be written by the Fock basis $\{|n\rangle\}$.  
A rank $k$ projection operator can always be written in the form 
\begin{equation}
U(\ |0\rangle\langle 0|+|1\rangle\langle 1|+\cdots 
+|k-1\rangle\langle k-1|\ )U^\dagger.  
\end{equation}
A diagonal projection operator $|n\rangle\langle n|$ corresponds to 
a radially symmetric configuration,  because it has the same number of 
creation and annihilation operators 
and $a^\dagger a\approx r^2/2$, where $r^2=x^2+y^2$.  
The most basic solution is $\lambda|0\rangle\langle0|$,  
which corresponds to the 
Gaussian lump configuration centered at the origin. 
More generally,  a field configuration $\phi_n(x,y)$ 
which corresponds to $|n\rangle\langle n|$ is 
\begin{equation}
\phi_n(x,y)=2(-1)^n e^{-r^2}L_n(2r^2),  
\end{equation}
where $L_n$ is the $n$-th Laguerre polynomial.  
If $U=1$,  the projection operator 
corresponds to a radially symmetric configuration 
centered at the origin whose width is $\sim \sqrt{n}$.  
If $U$ is a translation operator;  
$U(z)\equiv e^{za^\dagger-\bar{z} a}$ where $z={1\over\sqrt{2}}(x+iy)$,  
the corresponding field configuration has the same profile 
but centered at $(x,y)$ \cite{GMS}.  
A rank $k$ projection operator is called a level $k$ soliton in \cite{GMS},  
but in this letter we define a level $k$ soliton at $z$
as a projection operator onto a subspace spanned by 
$\{U(z)|0\rangle,U(z)|1\rangle,\dots,U(z)|k-1\rangle\}$.  
This can be shown to be equivalent to $k$ coincident level 1 solitons 
by the coordinate transformation in \cite{GHS}.  

The solution for a system of $k$ level 1 solitons \cite{GHS,HLRU} 
each centered at $z_i\ (i=1,\dots,k)$ 
on the complex $z$-plane can be constructed 
using coherent states 
\begin{equation}\label{defcoherent}
|z_i\rangle\equiv U(z_i)|0\rangle
=e^{-\frac12|z_i|^2}e^{z_ia^\dagger}|0\rangle,  
\quad U(z_i)=e^{z_i a^\dagger-\bar{z}_i a},  
\end{equation}
and is given as 
\begin{equation}\label{ksol}
\Phi=\lambda|z_i\rangle G^{ij}\langle z_j|,  
\end{equation}
or equivalently,  
\begin{equation}
\phi(z)=\lambda\cdot 
2G^{ij}G_{ji}e^{-2(\bar{z}-\bar{z}_j)(z-z_i)}.  
\end{equation}
Here $G_{ij}$ is the $n\times n$ hermitian matrix 
\begin{equation}
G_{ij}=\langle z_i|z_j\rangle
=e^{-\frac 12 |z_i|^2-\frac 12 |z_j|^2+\bar{z}_i z_j},  
\end{equation}
and $G^{ij}$ is its inverse such that $G^{il}G_{lj}=\delta^i_j$.  
The moduli space of this solution is parametrized by $z_i$.  
Its metric can be obtained by 
\begin{equation}
g_{i\bar{\jmath}}={1\over 2\pi\lambda^2}
\int d^2x \partial_i\phi\partial_{\bar{\jmath}}\phi
={1\over\lambda^2}
{\rm Tr}[\partial_i\Phi\partial_{\bar{\jmath}}\Phi] \label{physmetric}
\end{equation}
which comes from the time derivative term in the action 
when we regard $z_i$ depending (slowly) on time $t$
\cite{Manton,LRU}.  
The metric (\ref{physmetric}) may also be expressed as 
\begin{equation}
g_{i\bar{\jmath}}=G^{ij}\langle z_j|a\phi_\perp a^\dagger|z_i\rangle,  
\end{equation}
where $\phi_\perp\equiv 1-|z_l\rangle G^{lm}\langle z_m|$ and there is 
no summation over $i$ and $j$.  
This moduli space has a K\"ahler structure 
and the K\"ahler potential is given by the formula
\begin{eqnarray}\label{ksolKaehler}
K=\ln\left(\exp\left(\sum_{l=1}^k|z_l|^2\right)\cdot\det G\right)
=\ln\det(e^{\bar{z}_i z_j}).  
\end{eqnarray}
This K\"ahler potential has coordinate singularities when 
two or more $z_i$'s coincide.  
These singularities of K\"ahler potential would appear 
as conical singularities of the metric.  
These conical singularities 
have been explicitly examined in the case of $k=2$ and $k=3$ 
at the origin of the respective moduli space.  
For example,  in the case of $k=2$ ($z_1\ne z_2$),  
the metric is given as \cite{LRU} 
\begin{eqnarray}
&&d^2s=\frac 12f(r)(dr^2+r^2d\theta^2),  \nonumber\\
&&f(r)=\coth(r^2/4)-{r^2/4\over \sinh^2(r^2/4)},  \label{11metric}
\end {eqnarray}
where we have taken $z_1=-z_2=r e^{i\theta}/2\sqrt{2}$ 
so that the relative distance between two level 1 solitons is $r$.  
$f(r)$ behaves $\sim r^2$ as $r\rightarrow0$,  so if we take new 
coordinates $\rho=r^2$ and $\tilde{\theta}=2\theta$,  
the metric becomes a flat one near the origin:  
$ds^2\propto d\rho^2+\rho^2d\tilde{\theta}^2$.  
Thus a soliton coming from $\tilde{\theta}=0$ will pass through 
the origin (and the other soliton) 
smoothly and go in the direction of $\tilde{\theta}=\pi$,  
i.e.  $\theta=\pi/2$.  
That is,  the scattering of two level 1 solitons occurs 
at right angles.  
It is difficult to see the conical singularities 
explicitly in the K\"ahler metric (or potential) 
and to determine the deficit angle for higher $k$ or for a solution 
which contains higher level solitons,  
because the K\"ahler potential and metric is so complicated.  
The only exception is the case of scattering of two level $n$ solitons 
and it was conjectured in \cite{LRU} that 
the scattering occurs also at right angles.  

Now we will explore the locus of coincidence 
in the moduli space of a $(n,n')$-system 
by expanding a K\"ahler metric around the origin of the moduli space of 
a multi-soliton solution which consists of $k$ level 1 solitons.  

A K\"ahler potential for the moduli space of the $k$ soliton solution 
is given by (\ref{ksolKaehler}).  
Using the translational invariance of the theory,  
we can set the center of mass position $c=\frac 1k\sum_{i=1}^kz_i$ 
simply at the origin.  
Let $y_i$ be the relative coordinates $z_i-c$,  
then we have 
\begin{equation}
K=\ln\det(e^{\bar{y}_iy_j}).  
\end{equation}
Expanding the exponentials in the determinant,  we obtain 
\begin{equation}
\det(e^{\bar{y}_i y_j})={1\over k!}\sum_{m_1,\dots,m_k=0}^\infty 
{1\over m_1!\dots m_k!}|F_m(y)|^2,  
\end{equation}
where
\begin{equation}
F_m(y)\equiv \left|
\matrix{y_1^{m_1}&\cdots&y_k^{m_1}\cr
\vdots&\ddots&\vdots\cr y_1^{m_k}&\cdots&y_k^{m_k}} \right|,  
\quad m=(m_1,\dots,m_k).  
\end{equation}
For $\delta=(0,1,\dots,k-1)$,  
$F_\delta(y)$ becomes the Vandermonde determinant 
\begin{equation}\label{Vandermondedet}
F_\delta(y)=\prod_{i>j}(y_i-y_j).  
\end{equation}
$F_m(y)$ is non-zero for $0\le m_1<m_2<\dots<m_k$.  

Let $\mu=m-\delta=(\mu_1,\dots,\mu_k)$ 
($0\le \mu_1\le \mu_2\le \dots\le\mu_k$).  
Then $F_m(y)$ can be expressed as 
\begin{equation}
F_m(y)=S_\mu(y)F_\delta(y), 
\end{equation}
where $S_\mu(y)$ is a symmetric polynomial in $y_j$ 
of degree $\sum_{i=1}^k \mu_i$ and known as the Schur function \cite{Weyl}.  
It is defined for a partition of $\sum_i\mu_i$.  
So we may specify a Schur function with a partition 
instead of $\mu$:  for example,  
$S_{(0,0,\dots,0,1)}=S_1$,  
$S_{(0,\dots,0,1,1,2)}=S_{2,1,1}$,  etc.  
Their explicit examples are given as follows:  
\begin{eqnarray*}
&&S_1=e_1,  \\
&&S_2=e_1^2-e_2,\ 
S_{1,1}=e_2,  \\
&&S_3=e_1^3-2e_1e_2+e_3,  \ 
S_{2,1}=e_1e_2-e_3, \ 
S_{1,1,1}=e_3,  \\
&&S_4=e_1^4-3e_2e_1^2+2e_3e_1+e_2^2-e_4,  \ 
S_{3,1}=e_2e_1^2-e_2^2-e_3e_1+e_4,  \\
&&S_{2,2}=e_2^2-e_3e_1, \ 
S_{2,1,1}=e_3e_1-e_4,  \ 
S_{1,1,1,1}=e_4,  
\end{eqnarray*}
where $e_m$ denotes the $m$-th elementary symmetric polynomial:  
\begin{equation}
e_m=\sum_{i_{1}<\cdots <i_{m}}y_{i_{1}}\cdots y_{i_{m}},\quad
\end{equation}
Suppose that all of $y_i$ are small.  
the K\"ahler potential $K=\ln\det(e^{\bar{y}_i y_j})$ can be rewritten as  
\begin{eqnarray}
K&=&
\ln\left(\sum_{0\le\mu_1\le\cdots\le\mu_k}
{1\over \mu_1!(\mu_2+1)!\dots(\mu_k+k-1)!}
|S_\mu(y)|^2|F_\delta(y)|^2\right)\nonumber\\
&=&\ln\left\{
\left(1+\frac 1k|S_1|^2+{1\over k(k+1)}|S_2|^2+{1\over (k-1)k}|S_{1,1}|^2
+\dots\right)|F_\delta(y)|^2\right\}\nonumber\\
&&{}-\ln(1!\dots(k-1)!)\nonumber\\
&=&\sum_{m=1}^\infty{(-1)^{m-1}\over m!}G^m+\ln\prod_{i<j}|y_i-y_j|^2
-\ln(1!\dots(k-1)!),  
\end{eqnarray}
where 
\begin{eqnarray}
G&=&{1\over k}|S_1|^2+{1\over k(k+1)}|S_2|^2+{1\over (k-1)k}|S_{1,1}|^2
\nonumber\\
&&{}+{1\over k(k+1)(k+2)}|S_3|^2+{1\over(k-1)k(k+1)}|S_{2,1}|^2
+{1\over (k-2)(k-1)k}|S_{1,1,1}|^2\nonumber\\
&&{}+{1\over k(k+1)(k+2)(k+3)}|S_4|^2+{1\over (k-1)k(k+1)(k+2)}|S_{3,1}|^2
\nonumber\\
&&{}+{1\over (k-1)k(k+1)}|S_{2,2}|^2+{1\over(k-2)(k-1)k(k+1)}|S_{2,1,1}|^2
\nonumber\\
&&{}+{1\over(k-3)(k-2)(k-1)k}|S_{1,1,1,1}|^2+\cdots.  
\end{eqnarray}
Because we are working in the center of mass system, 
we have $e_1=\sum_iy_i=0$,  $S_2=-e_2$,  $S_{1,1}=e_2$ and so on.  
Thus $G$ can be simplified.
Furthermore,  note that $e_m$ can be rewritten by $p_m=\sum_{i=1}^k y_i^m$,  
e.g. $e_1=p_1,\ e_2=-\frac12 p_2+\frac12 p_1^2$.  
Finally,  we get the following expression for $K$:  
\begin{eqnarray}\label{expand}
K&=&\sum_{i<j}\ln|y_i-y_j|^2-\ln(1!\dots(k-1)!)
+{1\over 2(k^2-1)}|p_2|^2+{k\over3(k^2-4)(k^2-1)}|p_3|^2\nonumber\\
&&{}+{2\over (k-1)k(k+2)(k+3)}\left|\frac18p_2^2+\frac14p_4\right|^2
-{k^2+1\over 16(k^2-1)^2k^2}|p_2|^4\nonumber\\
&&{}+{2\over (k-3)(k-2)k(k+1)}\left|\frac18p_2^2-\frac14p_4\right|^2
+\cdots.  
\end{eqnarray}

The first term in the above expression 
will diverge when some of $y_i$ are coincident,  
but this divergence can be removed 
by the Jacobian of the coordinate transformation \cite{GHS}.  
Let us briefly recall how this can be achieved.  
For instance,  when $n$ of $z_i$ coincide at $z=c+w$, 
we should redefine the basis 
from $\{e^{z_ia^\dagger}|0\rangle\}_{i=1,\dots,k}$ to 
\begin{equation}
\left\{\sum_{a=1}^n(V^{-1})_{ab}e^{z_aa^\dagger}|0\rangle
\right\}_{b=1,\dots,n}
\ {\rm and}\quad \{e^{z_ja^\dagger}|0\rangle\}_{j=n+1,\dots,k},  
\end{equation}
where $V_{ab}\equiv(z_a-z)^{b-1}=(y_a-w)^{b-1}$.  
The new basis vectors for $b=1,\dots,n$ become 
\begin{equation}
{(a^\dagger)^{b-1}\over (b-1)!}e^{za^\dagger}|0\rangle,  \quad(b=1,\dots,n)
\end{equation}
as $z_a-z=y_a-w\rightarrow0$.  
It corresponds to the merging process of $n$ level 1 solitons into 
a single level $n$ soliton.  
This coordinate transformation changes the K\"ahler potential from $K$ to 
\begin{equation}\label{regularizedK}
K'=K+\ln(\det(V^{-1})^*\det(V^{-1}))=K-\sum_{a<b}\ln|y_a-y_b|^2,  
\end{equation}
in this limit.  Here we have used eq.(\ref{Vandermondedet}).  
$K$ and $K'$ are equivalent and give the same metric.  
Thus the singular terms in $K$ could be removed 
by this coordinate transformation.  
This procedure is further applicable to the remaining $z_j\ (j=n+1,\dots,k)$ 
repeatedly.  
Therefore,  we can get a expansion of a K\"ahler potential 
corresponding to an arbitrary level-$(n_1,n_2,\dots)$ soliton solution 
($n_1+n_2+\cdots=k$) in the locus of coincidence from eq.(\ref{expand}).

Now we can explore how the scattering of one level $n$ soliton 
and one level $n'$ soliton is ($n+n'=k$).  
We call this system the $(n,n')$-system.  
We will take $y_1=\cdots=y_n=n'y/k,\ y_{n+1}=\cdots=y_k=-ny/k$,  
such that the relative distance is always $r=\sqrt{2}|y|$.  
Then eq.(\ref{regularizedK}) becomes 
\begin{eqnarray}
K'&=&{n^2n'^2\over 2(k^2-1)k^2}|y|^4
+{n^2n'^2(n-n')^2\over 3(k^2-4)(k^2-1)k^3}|y|^6\nonumber\\
&&{}+{n^2n'^2c_8(k,n)\over 4(k^2-9)(k^2-4)(k^2-1)^2k^4}|y|^8+\cdots,  
\end{eqnarray}
where,  
\begin{equation}
c_8(k,n)=k^6-10k^5n+34k^4n^2+10k^3n-k^2-46k^2n^2+25k^2m^4+72kn^3-36k^4.  
\end{equation}
Then we get the K\"ahler metric by 
$g_{y\bar{y}}=\partial_y\partial_{\bar{y}}K'$;  
\begin{eqnarray}
ds^2&=&\left({n^2n'^2\over 2(k^2-1)k^2}r^2+\cdots\right)
(dr^2+r^2d\theta^2),  \quad y={1\over\sqrt{2}}re^{i\theta}\label{genexpand}
\end{eqnarray}
This behavior is the same as the case of the (1,1)-system 
as we have seen in eq.(\ref{11metric}).  
So we conclude that the scattering of the $(n,n')$-system occurs 
at right angles in the center of mass system 
when the impact parameter is zero.  

This behavior of the $(n,n')$ scattering can be explicitly examined  
by the K\"ahler potential using new basis.   
If we take $k$ linearly independent (not necessarily orthogonal) 
basis vectors $\{|\psi_i\rangle\}$ which span a subspace of ${\cal H}$,  
the rank $k$ projection operator onto this subspace is given by 
\cite{GHS,HLRU}
\begin{equation}
P=|\psi_i\rangle h^{ij}\langle\psi_j|.    
\end{equation}
Here,  $h^{ij}$ is the inverse matrix of the $k\times k$ hermitian matrix 
\begin{equation}
h_{ij}\equiv\langle\psi_i|\psi_j\rangle,  
\end{equation}
so that $h^{ik}h_{kj}=\delta^i_j$.  
If the $k$ basis vectors $\{|\psi_i\rangle\}$ holomorphically depend on 
complex parameters $z_a$,    
a natural metric on the moduli space is  
\begin{equation}
g_{a\bar{b}}={\rm Tr}[\partial_aP\partial_{\bar{b}}P],  
\end{equation}
and a K\"ahler potential which gives this metric is given by \cite{GHS}
\footnote{For the $k$ level 1 solitons, 
if we define $|\psi_i\rangle\equiv e^{z_ia^\dagger}|0\rangle$ 
($i=1,\dots,k$) as in ref.\cite{GHS} which is differ from $|z_i\rangle$ 
in eq.(\ref{defcoherent}) by the normalization factor, 
then eq.(\ref{genKaehler}) gives the same result 
as eq.(\ref{ksolKaehler}).  }
\begin{equation}\label{genKaehler}
K=\ln\det(h_{ij}).  
\end{equation}
The above metric is the same one 
as the physical metric 
\begin{equation}
g_{a\bar{b}}
={1\over 2\pi\lambda^2}\int d^2x \partial_a\phi\partial_{\bar{b}}\phi.  
\end{equation}
up to a overall constant.  

Now take the $k=n+n'$ basis vectors as 
\begin{eqnarray}
\{|\psi_i\rangle\}&\equiv&
\left\{ e^{w_1a^\dagger}|0\rangle,\ a^\dagger e^{w_1a^\dagger}|0\rangle,\ 
\dots,\  {a^\dagger}^{n-1} e^{w_1a^\dagger}|0\rangle,\ \right.\nonumber\\
&&\qquad \left.e^{w_2a^\dagger}|0\rangle,\ 
a^\dagger e^{w_2a^\dagger}|0\rangle,\ 
\dots,\ {a^\dagger}^{n'-1}e^{w_2a^\dagger}|0\rangle\right\}  \nonumber\\
&=&\left\{{a^\dagger}^{n_i}e^{z_ia^\dagger}|0\rangle\right\}_{i=1,\dots,k},  
\label{basis1}
\end{eqnarray}
with 
\begin{eqnarray}
&&z_1=\cdots=z_n=w_1,\quad  z_{n+1}=\cdots=z_k=w_2,  \nonumber\\
&&n_i=\cases{i-1,  &$1\le i\le n$,  \cr 
i-n-1,  &$n+1\le i\le k$.  }\label{zini}
\end{eqnarray}
For this basis,  the inner product matrix $h_{ij}$ is expressed as follows:  
\begin{equation}
h_{ij}=\langle \psi_i|\psi_j\rangle
=\cases{n_i!\bar{z}_i^{n_j-n_i}L_{n_i}^{(n_j-n_i)}(-\bar{z}_iz_j)
\cdot e^{\bar{z}_iz_j},  
&$n_i\le n_j$,   \cr
n_j!z_j^{n_i-n_j}L_{n_j}^{(n_i-n_j)}(-\bar{z}_iz_j)\cdot e^{\bar{z}_iz_j},  &
$n_j\le n_i$.  }
\end{equation}
Here,  $L_n^{(\alpha)}$ is the Laguerre polynomial:  
\begin{equation}\label{Laguerre}
L_n^{(\alpha)}(x)=\sum_{r=0}^n(-1)^r{n+\alpha\choose n-r}{x^r\over r!}.
\end{equation}

\begin{figure}
\hspace*{3.5cm}
\includegraphics{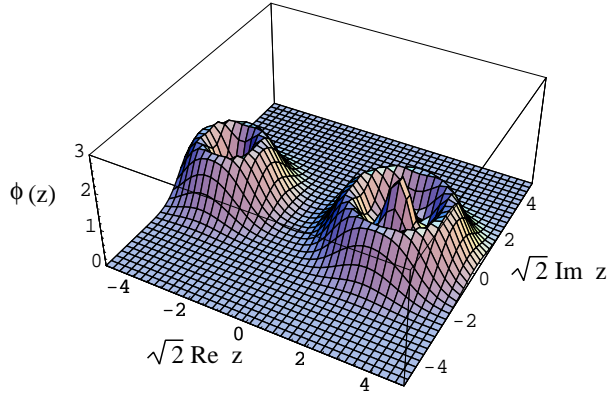}
\caption{
An example of the $(n,n')$-systems.  This is a (3,2)-system 
plotted by using the exact solution for a $(n,n')$-system in case of $r=6$.  }
\label{32sys}
\end{figure}

{}From the Weyl-Moyal correspondence,  we obtain the solution of 
the $(n,n')$-system as 
\begin{equation}
\phi(z)=\lambda h^{ij}\phi_{ij}(z),  
\end{equation}
where $\phi_{ij}(z)$ is 
a field configuration corresponding to $|\psi_i\rangle\langle\psi_j|$ .
This is given by the Weyl-Moyal correspondence:  
\begin{eqnarray}
&\phi_{ij}(z)&=\cases{
(-)^{n_i} n_i!(2z-z_i)^{n_j-n_i}
L_{n_i}^{(n_j-n_i)}((2\bar{z}-\bar{z}_j)(2z-z_i))\cdot{\phi_0}_{ij}, 
&$n_i\le n_j$ \cr
(-)^{n_j} n_j!(2\bar{z}-\bar{z}_j)^{n_i-n_j}
L_{n_j}^{(n_i-n_j)}((2\bar{z}-\bar{z}_j)(2z-z_i))\cdot{\phi_0}_{ij}, 
&$n_j\le n_i$,  }
\end{eqnarray}
Here,  ${\phi_0}_{ij}(z)\equiv 2e^{\bar{z}_jz_i}
e^{-2(\bar{z}-\bar{z}_j)(z-z_i)}$.  
Fig.\ref{32sys} is an example of a field configuration 
for a $(n,n')$-system plotted by using the above formulae.  

In the case of the $(n,1)$-system,  the K\"ahler potential 
can be calculated as follows.  
Let us change the basis from $|\psi_i\rangle$ to 
\begin{equation}\label{basis2}
|\tilde{\psi}_i\rangle\equiv U(z_i)|n_i\rangle\ (i=1,\dots,k), 
\quad U(z_i)=e^{z_ia^\dagger-\bar{z}_i a},  
\end{equation}
where $|n_i\rangle$ are the Fock basis. 
$z_i$ and $n_i$ are the same as (\ref{zini}).  
The matrix $h_{ij}$ can be re-expressed with the $k\times k$ matrices 
$B_{ij}\equiv\langle\tilde{\psi}_i|\psi_j\rangle$ and 
$\tilde{h}_{ij}\equiv\langle\tilde{\psi}_i|\tilde{\psi}_j\rangle$ 
which have structures simpler than $h$;  
\begin{equation}\label{rewriteh}
h_{ij}=\langle \psi_i|\psi_j\rangle
=\langle \psi_i|\tilde{\psi}_l\rangle 
\tilde{h}^{lm}\langle \tilde{\psi}_m|\psi_j\rangle
=B^*_{il}\tilde{h}^{lm}B_{mj}=(B^*\tilde{h}^{-1} B)_{ij}.  
\end{equation}
Here,  we have introduced the inverse matrix $\tilde{h}^{ij}$ 
of $\tilde{h}_{ij}$ such that $\tilde{h}^{il}\tilde{h}_{lj}=\delta^i_j$.  
The second equality in eq.(\ref{rewriteh}) 
holds because $\{|\psi_i\rangle\}$ and $\{|\tilde{\psi}_i\rangle\}$ 
span the same subspace of ${\cal H}$  
and $|\tilde{\psi}_l\rangle \tilde{h}^{lm}\langle\tilde{\psi}_m|$ 
is the identity operator in this subspace.  
With the relation,  we have 
\begin{equation}\label{deth1}
\det h={\det B^* \det B\over \det \tilde{h}}.  
\end{equation}
{}From (\ref{basis1}) and (\ref{basis2}),  $B_{ij}$ is given by 
\begin{equation}
B_{ij}=\cases{
\sqrt{n_i!}e^{-\frac12|z_i|^2+\bar{z}_iz_j}\bar{z}_i^{n_j-n_i}
L_{n_i}^{(n_j-n_i)}(-\bar{z}_i(z_j-z_i)),  &$n_i\le n_j$,  \cr
{n_j!\over\sqrt{n_i!}}e^{-\frac12|z_i|^2+\bar{z}_iz_j}(z_j-z_i)^{n_i-n_j}
L_{n_j}^{(n_i-n_j)}(-\bar{z}_i(z_j-z_i)),  &$n_j\le n_i$. }
\end{equation}
For the $(n,1)$-system,  note that $z_i=z_j=w_1\ (i,j\le n)$ 
and $L_n^{(\alpha)}(0)={n+\alpha\choose n}$.  
Then we obtain 
\begin{equation}
B=\left(\matrix{
C_1&\bar{w}_1C_1&\bar{w}_1^2C_1&\cdots&\bar{w}_1^{n-1}C_1&C_2\cr
&{1\choose 1}C_1&{2\choose1}\bar{w}_1C_1&\cdots
& {n-1\choose1}\bar{w}_1^{n-2}C_1
	&(w_2-w_1)C_2\cr
&&\sqrt{2!}{2\choose2}C_1&\cdots&\sqrt{2!}{n-1\choose2}\bar{w}_1^{n-3}C_1
&{(w_2-w_1)^2\over\sqrt{2!}}C_2\cr
&\mbox{\Large 0}&&\ddots&\vdots&\vdots\cr
&&&&\sqrt{(n-1)!}{n-1\choose n-1}C_1
&{(w_2-w_1)^{n-1}\over\sqrt{(n-1)!}}C_2\cr
C_3&\bar{w}_2C_3&\bar{w}_2^2C_3&\cdots&\bar{w}_2^{n-1}C_3&C_4
}\right),  
\label{matrixB}
\end{equation}
where $C_1=e^{\frac12|w_1|^2},\ C_2=e^{-\frac12|w_1|^2+\bar{w}_1w_2},\ 
C_3=e^{-\frac12|w_2|^2+\bar{w}_2 w_1},\ C_4=e^{\frac12|w_2|^2}$.  
Noting $U(z_i)a^\dagger=(a^\dagger-\bar{z}_i)U(z_i)$,  $\tilde{h}_{ij}$ can 
be written for the $(n,1)$-system as 
\begin{equation}
\tilde{h}=\left(\matrix{
1&&&&&G_{12}\cr
&1&&\mbox{\Large 0}&&(w_2-w_1)G_{12}\cr
&&1&&&{(w_2-w_1)^2\over\sqrt{2!}}G_{12}\cr
&\mbox{\Large 0}&&\ddots&&\vdots\cr
&&&&1&{(w_2-w_1)^{n-1}\over\sqrt{(n-1)!}}G_{12}\cr
G_{21}&(\bar{w}_2-\bar{w}_1)G_{21}
&{(\bar{w}_2-\bar{w}_1)^2\over\sqrt{2!}}G_{21}&\cdots
&{(\bar{w}_2-\bar{w}_1)^{n-1}\over\sqrt{(n-1)!}}G_{21}&1}\right),  
\label{matrixhtilde}
\end{equation}
where $G_{12}=\langle w_1|w_2\rangle,\ G_{21}=\langle w_2|w_1\rangle$.  
{}From these matrix forms,  we can easily calculate their determinants:  
\begin{eqnarray}
\det B&=&\left(\prod_{m=0}^{n-1}\sqrt{m!}\right)e^{\frac n2|w_1|^2}
e^{\frac12|w_2|^2}\left(1-e^{-|w_1-w_2|^2}
\sum_{m=0}^{n-1}{|w_1-w_2|^{2m}\over m!}\right),  \label{detB}\\
\det\tilde{h}&=&1-e^{-|w_1-w_2|^2}\sum_{m=0}^{n-1}{|w_1-w_2|^{2m}\over m!}.  
\label{dethtilde}
\end{eqnarray}
{}From eqs.(\ref{deth1}),  (\ref{detB}) and (\ref{dethtilde}),  
the determinant of $h$ or 
the K\"ahler potential is determined as follows;  
\begin{eqnarray}
K&=&n|w_1|^2+|w_2|^2+\ln\left(1-e^{-|w_1-w_2|^2}
\sum_{m=0}^{n-1}{|w_1-w_2|^{2m}\over m!}\right)
+\ln\left(\prod_{m=0}^{n-1}m!\right)\nonumber\\
&=&\frac n{n+1} |y|^2+\ln\left(1-e^{-|y|^2}\sum_{m=0}^{n-1}{|y|^{2m}\over m!}
\right)+\ln\left(\prod_{m=0}^{n-1}m!\right).  \label{1nKaehler}
\end{eqnarray}
In the second line,  we have set $w_1=y/(n+1),\ w_2=-ny/(n+1)$.  
The K\"ahler metric can be calculated from this as 
\begin{equation}
g_{y\bar{y}}=\partial_y\partial_{\bar{y}}K=\frac n{n+1}
+{|y|^{2(n-1)}\over(n-1)!}{e^{-|y|^2}\over Q(|y|^2)}
\left(n-|y|^2-{|y|^{2n}\over(n-1)!}{e^{-|y|^2}\over Q(|y|^2)}\right),  
\end{equation}
where 
\begin{equation}
Q(x)\equiv 1-e^{-x}\sum_{m=0}^{n-1}{x^m\over m!}.  
\end{equation}
We introduce coordinates $(r,\theta)$ by $y=re^{i\theta}/\sqrt{2}$ as before.  
The metric becomes $ds^2=f(r)(dr^2+r^2d\theta^2)$,  where
\begin{equation}
f(r)\equiv\frac12\left\{\frac n{n+1}+{(r^2/2)^{n-1}\over(n-1)!}
{e^{-r^2/2}\over Q(r^2/2)}
\left(n-{r^2\over 2}-{(r^2/2)^n\over(n-1)!}
{e^{-r^2/2}\over Q(r^2/2)}\right)\right\}.  
\end{equation}
Expanding this around $r\sim0$,  we get 
\begin{equation}\label{n1approxds2}
ds^2\approx{n\over2(n+1)^2(n+2)}r^2
(dr^2+r^2d\theta^2).  
\end{equation}
This indeed agrees with the result (\ref{genexpand}).  
When the relative distance $r$ 
between the level $n$ soliton and the level 1 soliton is very large,  
$Q(r^2/2)$ goes to 1.  Immediate consequence of this fact is that 
the K\"ahler potential (\ref{1nKaehler}) behave as $K\approx n|y|^2/(n+1)$ 
and thus the metric goes to a flat one in this limit;  
\begin{equation}
ds^2\approx \frac n{2(n+1)}(dr^2+r^2d\theta^2).  
\end{equation}
That is,  two solitons do not affect each other when they are remote.  
This is the same as the case of the (1,1)-system.  

\begin{figure}
\hspace*{4cm}
\includegraphics{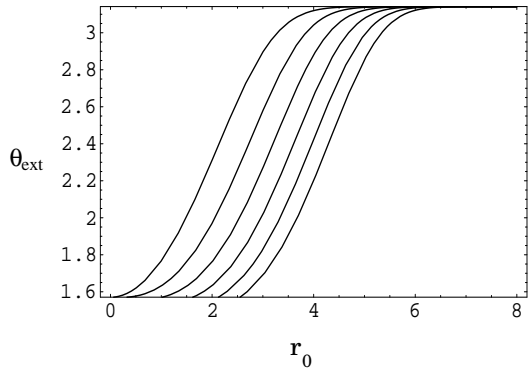}
\caption{The scattering angle for $(n,1)$-systems.  
The curves indicate the case of 
$n=1,2,3,4,5,6$ (from left to right in the graph) respectively.  }
\label{ext1-6}
\end{figure}

\begin{figure}
\hspace*{4cm}
\includegraphics{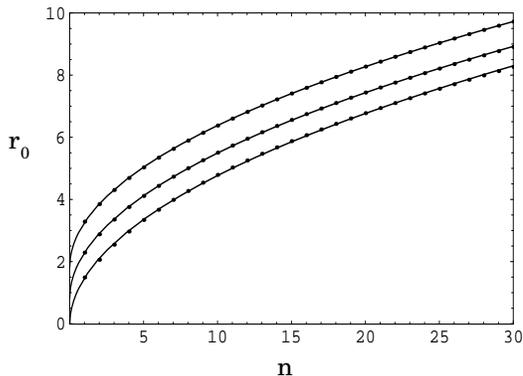}
\caption{The closest distance 
($n\le 30$) for $\theta_{\rm ext}=2.0,2.5,3.0$ 
(from bottom to top in the graph).  
The curves are the fitting curves in the text.  }
\label{sqrtn}
\end{figure}

Let us consider the $(n,1)$ scattering more.  
Using the above $f(r)$, we can calculate numerically the scattering angle 
in case that the impact parameter is nonzero.  
The scattering angle $\theta_{{\rm ext}}$ is given by the formula \cite{LRU} 
\begin{equation}\label{exitangle}
\theta_{{\rm ext}}=-2\int_\infty^{r_0}{ds\over s\sqrt{
\left(s^2f(s)\Big/ r_0^2f(r_0)\right)-1}},  
\end{equation}
where $r_0$ is the closest distance and related to the impact parameter $b$ by 
$b=r_0\sqrt{f(r_0)}$.  
For $n\le 6$, the scattering angles are plotted in fig.\ref{ext1-6}.   
When the closest distance goes to zero i.e. the zero impact parameter limit,  
a scattering angle is $\pi/2$.  
As the closest distance or an impact parameter becomes large,  
a scattering angle closes to $\pi$ (no scattering limit).  
This figure shows that for a fixed $\theta_{\rm ext}$ 
the closest distance $r_0$ seems to be roughly proportional to $\sqrt{n}$.  
This is expected from the fact that the (radially symmetric) level $n$ soliton 
has size $\sqrt{n}$.
In fact,  when $\theta_{\rm ext}$ is fixed to be 2.0,  
we observe that $r_0({\theta_{\rm ext}=2.0})\approx -0.05+1.52\sqrt{n}$.  
Similarly,  when $\theta_{\rm ext}$ is fixed to be 2.5 and 3.0,  
$r_0({\theta_{\rm ext}=2.5})\approx 0.81+1.48\sqrt{n}$ 
and $r_0({\theta_{\rm ext}=3.0})\approx 1.81+1.44\sqrt{n}$ 
respectively.  These fittings are drawn in fig.\ref{sqrtn}. 
These observations are natural reflection of the width  of the soliton, 
though it is difficult to see this from the formula (\ref{exitangle}).  

In this letter,  we have seen some properties of a general $(n,n')$-system.  
We conclude that the $(n,n')$ scattering occurs at right angles 
in the case of the zero impact parameter.  
We have confirmed that the right angle scattering is a universal property 
of two-body scattering of noncommutative solitons.  
Especially for the $(n,1)$-system,  we have shown directly 
the metric of the relative moduli space goes to a flat one 
far from the origin.  
This could be generalized to arbitrary $(n,n')$-systems.  
Finally we have numerically calculated the scattering angle 
for the $(n,1)$-system 
and found the closest distance for a fixed scattering angle 
is well approximated by a function 
$a+b\sqrt{n}$ ($a$ and $b$ are some numerical constants).  
It may exist the similar relation for a $(n,n')$ scattering.  

Note that as $n$ becomes large,  the K\"ahler potential (\ref{1nKaehler}) 
for the $(n,1)$-system diverges,  since $Q(x)\rightarrow 0$.  
But even in this case,  expanding the potential in terms of $|y|^2/n$,  
and neglecting a coordinate singularity $\ln|y|^2$,  
we can get the same formula for the metric as eq.(\ref{n1approxds2}).  
The metric tends to be zero everywhere when $n$ goes to infinity.  
If we rescale $y\rightarrow \sqrt{n}y$,  then we see that 
in this new coordinate the scattering behavior of this system is 
qualitatively the same as the case of finite $n$.  

It is interesting to consider corrections to the metric 
when the noncommutativity parameter is large but finite.  
It is also interesting to explore the multi soliton solutions 
and their properties when gauge degrees of freedom are exist,  
where the solutions we have considered in this letter can be 
a part of exact solutions even at finite $\theta$\cite{HKL}.  
These subjects are left for future studies.  

{\bf Acknowledgment}
The research of KI is supported in part by Grant-in-Aid from the
Ministry of Education, Science, Sports and Culture of Japan, Priority
Area 707 ``Supersymmetry and Unified Theory of Elementary Particles''.

\end{document}